\documentclass[twocolumn,aps,prl,superscriptaddress,showpacs,floatfix]{revtex4}

\usepackage{epsfig,bm,feynmf}

\usepackage{graphics}

\usepackage[normalem]{ulem} 
\renewcommand\sout{\bgroup \color{red} \ULdepth=-.5ex \ULset}
\usepackage[dvips]{color} 

\def\ltap{\ \raise.3ex\hbox{$<$\kern-.75em\lower1ex\hbox{$\sim$}}\ }
\def\gtap{\ \raise.3ex\hbox{$>$\kern-.75em\lower1ex\hbox{$\sim$}}\ }

\begin{document}



\title{$\Lambda_{c}$ enhancement from
strongly coupled quark-gluon plasma}

\author{Su Houng Lee}
\affiliation{Institute of Physics and Applied Physics, Yonsei
University, Seoul 120-749, Republic of Korea}
\author{Kazuaki Ohnishi}
\affiliation{Institute of Physics and Applied Physics, Yonsei
University, Seoul 120-749, Republic of Korea}
\author{Shigehiro Yasui}
\affiliation{Institute of Physics and Applied Physics, Yonsei
University, Seoul 120-749, Republic of Korea}
\affiliation{Department of Physics, National Taiwan University, Taipei 10617, Taiwan}
\author{In-Kwon Yoo}
\affiliation{Pusan National University, Pusan 609-735, Republic of
Korea}
\author{Che Ming Ko}
\affiliation{Cyclotron Institute and Physics Department, Texas A\&M
University, College Station, TX 77843, U.S.A.}


\begin{abstract}
We propose the enhancement of $\Lambda_{c}$ as a novel QGP signal in
heavy ion collisions at RHIC and LHC.  Assuming a stable bound
diquark state in the strongly coupled QGP
near the critical temperature, we argue that the direct
two-body collision between a $c$ quark and a $[ud]$
diquark would lead to an enhanced $\Lambda_{c}$ production in
comparison
 with the normal three-body collision among independent
$c$, $u$ and $d$ quarks. In the coalescence model, we find that the
$\Lambda_{c}/D$ yield ratio is enhanced substantially due to the
diquark correlation.
\end{abstract}



\maketitle

The quark-gluon plasma (QGP) is one of the most actively pursued
subjects in strong interaction physics. Recent experimental and
theoretical studies have revealed intriguing properties of QGP.
These range from the realization of the perfect fluid behavior
\cite{Teaney:2000cw} and consequently of the strong coupling nature
of QGP \cite{Policastro:2001yc} to the lattice findings
suggesting that the $c\bar{c}$ bound state could survive up to
temperatures well above the critical temperature $T_{\rm C}$
\cite{Datta:2002ck, Asakawa:2003re}. Thus a new picture of QGP with
non-trivial correlations has emerged, and this state is nowadays
called the strongly coupled QGP (sQGP) \cite{Shuryak:2003ty,
Asakawa:2003re, Hatsuda:1985eb}.

Although the lattice QCD indicates the absence of $q\bar s$ bound
states in QGP \cite{Koch:2005vg}, the color non-singlet $qq$, $qg$,
$gg$ bound states may exist if there are attractive channels between
the constituents \cite{Shuryak:2003ty}. Among them, the diquark $qq$
with color multiplets $\bar{\bf 3}_{\rm c}$ and ${\bf 6}_{\rm c}$ is
the simplest one \cite{Jaffe:1976ig}. In flavor SU(3)$_{\rm f}$, the
$\bar{\bf 3}_{\rm c}$ diquarks are classified into a scalar with
$\bar{\bf 3}_{\rm f}$ (``good" diquark) and a vector with $\bf
6_{\rm f}$ (``bad" diquark) \cite{Jaffe:1976ig}, which are in the
attractive and repulsive channels, respectively, according to the
one-gluon exchange (OGE) picture. Concerning ${\bf 6}_{\rm c}$, the
diquarks belong to either the scalar with $\bf 6_{\rm f}$ or the
vector with $\bar{\bf 3}_{\rm f}$. Although the latter is in an
attractive channel, its strength is only one-sixth of that in the
attractive $\bar{\bf 3}_{\rm c}$ state according to the OGE. Recent
lattice calculations support the diquark correlation in vacuum
\cite{Alexandrou:2006cq}.

The diquark not only has relevance to the color superconductivity at
high density \cite{Alford:1997zt} and to the Bose-Einstein
Condensate (BEC) in the strong coupling regime \cite{Nishida:2005ds}
but also is an interesting object for understanding heavy baryons
$Qqq$ with one heavy quark $Q=c$, $b$ and two light quarks $q=u$,
$d$ \cite{Jaffe:2004ph}. Generally, in the heavy quark mass limit,
the light quarks are almost decoupled from the heavy quark
\cite{Jaffe:2004ph}, as the $Qq$ color-spin interaction is
suppressed in OGE by the heavy quark mass \cite{De Rujula:1975ge}
and in the instanton model by the small coupling between the heavy
quark and the instanton vacuum as a result of the absence of
zero-energy modes \cite{Oka:1989ud}. Therefore, $Qqq$ baryons only
have a strong correlation within the light quark sector as in the
conventional diquark model \cite{GellMann:1964nj,
Lichtenberg:1982jp, Ebert:1995fp, Semay:1994ht, Santopinto:2004hw}
and in models that treat the triquark $q\bar{q}\bar{q}$ in exotic
$D_s$ mesons as a color non-singlet state
\cite{Jaffe:2004ph,Yasui:2007dv}. In this viewpoint, $\Lambda_{c}$
($\Lambda_{b}$) can be regarded as an ideal two-body system composed
of the $c$ ($b$) quark and the $[ud]$ diquark, {\it i.e.}, $c[ud]$
($b[ud]$). In contrast, the $\Lambda$ with an $s$ quark may not be
such a simple quark-diquark system, because SU(3)$_{\rm f}$ symmetry
allows interactions among the three quarks.

In heavy ion collisions, open and hidden charmed hadrons are
interesting observables for studying the QGP \cite{Matsui:1986dk},
particularly at LHC as an appreciable number of $c\bar{c}$ pairs is
expected to be produced. Moreover, the ALICE detector at LHC is
designed to measure charmed particles with enhanced vertex tracking
system, which has a spatial resolution of 12 $\mu$m with the best
precision \cite{alice1}. For the planned upgrade of STAR and PHENIX
detectors at RHIC, additional vertex detectors will be added to
achieve direct vertex reconstruction of charmed particles
\cite{STAR1,PHENIX1}. With such precise measurement of the vertices,
even the measurement of open and hidden bottomed hadrons is
possible.

The existence of diquark correlations in QGP can be probed by
studying their effects on $\Lambda_c$ production in relativistic
heavy ion collisions \cite{sateesh}. One of the important findings RHIC is that a
new hadronization mechanism, based on the coalescence of constituent
quarks, is operative in heavy ion collisions
\cite{biro95,Voloshin03,Hwa:2002tu,greco,Fries:2003vb}. Here,
instead of fragmentation, hadronization takes place by the
recombination of partons in QGP or by their collisions into final
hadrons. The coalescence model has been quite successful in
describing the pion and proton transverse momentum spectra at
intermediate momenta as well as at low momenta if resonances are
included \cite{greco1}. It also gives a natural account for the
observed constituent quark number scaling of hadron elliptic flows
\cite{kolb} and the large elliptic flow of charmed mesons
\cite{greco2}. In such a picture, the $\Lambda_{c}$ is formed from
the three-body collisions among the $c$, $u$ and $d$ quarks at the
critical temperature of QGP.  If there are strong diquark
correlations in QGP at this temperature, then $\Lambda_{c}$ could be
additionally formed from the two-body collisions between the $c$
quark and the $[ud]$ diquark. Here, the diquark structure in
$\Lambda_{c}$ is essential to the direct two-body production of
$\Lambda_{c}$, because additional process is needed to break up the
diquark correlation if the diquark is absent inside $\Lambda_{c}$.
Since the two-body collision generally dominates over multi-body
collisions we thus expect an enhanced $\Lambda_{c}$ yield in heavy
ion collisions if there are diquark correlations, and this could be
a new signal for the search of the QGP.

The binding energy of the lightest scalar $\bar{\bf 3}_{\rm f}$ and
$\bar{\bf 3}_{\rm c}$ $[ud]$ diquark can be estimated using a
simplified constituent quark model based on the color-spin
interaction. This model has been shown to describe very well the
mass differences between various hadrons including the charmed ones
\cite{Lee:2007tn}. In vacuum, the color-spin interaction between two
quarks gives the $[ud]$ diquark mass as $m_{\rm[ud]} = m_{\rm u} +
m_{\rm d} - C \vec{s}_{\rm u} \cdot \vec{s}_{\rm d} / m_{\rm
u}m_{\rm d} $ with the quark mass $m_{\rm u}=m_{\rm d}=0.3$ GeV, the
spin operator $\vec{s}_{i}$ ($i={\rm u}, \rm{d}$), and a constant
$C/m_{\rm u}^{2}=0.193$ GeV  fitted to the $N - \Delta$ splitting
\cite{Lee:2007tn}. The color-spin interaction effectively contains
the non-perturbative dynamics in vacuum, and hence gives the maximum
binding energy 0.145 GeV of the diquark in the strong coupling
limit. The diquark mass is, however, expected to increase in QGP
where the coupling would become smaller than that in vacuum. In the
analysis of Ref.\cite{Shuryak:2003ty}, the zero binding of diquark
occurs slightly above $T_c$. Since the strength of the color-spin
interaction is of the same order as the critical temperature $T_{\rm
C} \simeq 0.17$ GeV, the diquark correlation could still be present
near $T_{\rm C}$. Therefore, we use the diquark mass ranging from
$m_{\rm[ud]}=0.455$ GeV for the maximum binding to $m_{\rm[ud]}=0.6$
GeV for the threshold.

For the dynamics of heavy ion collisions, we follow the expanding
fire-cylinder model, which leads to the volume $V_{\rm C}\simeq
1000$ ${\rm fm}^3$ in central Au+Au collisions at $\sqrt{s_{\rm
NN}}=200~{\rm GeV}$ \cite{chen} and $V_{\rm C} \simeq 2700$ ${\rm
fm}^3$ in central Pb+Pb collisions at $\sqrt{s_{\rm NN}}=5.5~{\rm
TeV}$ \cite{Zhang:2007dm}.  At $T_{\rm C}=0.175~{\rm GeV}$, the
equilibrium light quark numbers in QGP are $N_{\rm u}=N_{\rm d}
\simeq 245$ \cite{chen} and 662 \cite{Zhang:2007dm} in collisions at
RHIC and LHC, respectively, all in one unit of midrapidity. The
equilibrium diquark numbers at RHIC and LHC for this temperature are
estimated as $N_{\rm[ud]}\simeq 77$ and $208$, respectively, for
$m_{\rm[ud]}=0.455$ GeV, and $N_{\rm[ud]}\simeq 44$  and $119$,
respectively, for $m_{\rm[ud]}=0.6$ GeV. For the charm quark number
at the phase transition temperature, we take it to be the same as
that produced from the initial hard scattering of nucleons in the
colliding nuclei, and their numbers in one unit of midrapidity are $N_{\rm c}\simeq 3$ and 20,
respectively, at RHIC \cite{chen} and LHC \cite{Zhang:2007dm}. We
thus neglect charm production from QGP, which is unimportant at RHIC
but could be significant at LHC if the initial temperature of QGP is
high \cite{Zhang:2007dm}. The charm quarks are assumed to reach
thermal equilibrium in QGP, and this is consistent with the observed
large elliptic flow of the electrons from the decay of charmed
mesons in heavy ion collisions at RHIC
\cite{Adler:2005ab,Laue:2004tf}, which requires that charm quarks
interact strongly in QGP and are thus likely to reach thermal
equilibrium \cite{van Hees:2004gq,Zhang:2005ni,Molnar:2004ph}.

For coalescence of $c$ quarks with independent or uncorrelated $u$
and $d$ quarks in QGP, the contribution to the number of produced
$\Lambda_c$ is given by \cite{Chen:2003tn,chen}
\begin{eqnarray}\label{coalmodel}
N_{\Lambda_{c}({\rm cud})}^{\rm coal} &=& g_{\Lambda_{c}({\rm
cud})}\int_{\sigma_{\rm C}} \prod_{i=1}^{n=3}\frac{p_{i}\cdot {\rm
d}\sigma _{i}{\rm d}^{3}\mathbf{p}_{i}}{(2\pi )^{3}E_{i}}f_{\rm
q}(x_{i},p_{i}) \label{coal} \nonumber\\
&& \times f_{\Lambda_c}^{\rm W}(x_{1}..x_{n};p_{1}..p_{n}),
\end{eqnarray}
where $g_{\Lambda_{c}({\rm cud})}=2\times1/3^3\times1/2^3=1/108$ is
the color-spin-isospin factor for the three quarks to form
$\Lambda_{c}$, and ${\rm d}\sigma$ denotes an element of a
space-like hypersurface of QGP at hadronization. Following Ref.\
\cite{chen}, we adopt the $u$ and $d$ as well as the $c$ quark
momentum distribution function $f_{\rm q}(x,p)$ with Bjorken
correlation between the space-time rapidity and the momentum-energy
rapidity, and the $\Lambda_{c}$ Wigner distribution function
$\label{wigner1} f_{\Lambda_{c}({\rm cud})}^{\rm W}(x;p)=8^2\exp
(-\sum_{i=1}^{2}{\mathbf{y}_{i}^{2}}/{\sigma_{i}^{2}}
-\sum_{i=1}^{2}\mathbf{k}_{i}^{2}\sigma_{i}^{2}),$ where the
relative coordinates $\mathbf{y}_{i}$ and momenta $\mathbf{k}_{i}$
are related to the quark coordinates $\mathbf{x}_{i}$ and momenta
$\mathbf{p}_{i}$ by the Jacobian transformations defined in Eqs.~(7)
and (8) of Ref.~\cite{chen}. The width parameter $\sigma_i$ in the
Wigner function is related to the oscillator frequency $\omega$ by
$\sigma_i=1/\sqrt{\mu_i\omega}$ with the reduced masses $\mu_i$
defined in Eq.~(9) of Ref.~\cite{chen}. Neglecting the transverse
flow as well as using the non-relativistic approximation, we obtain
\cite{chen}:
\begin{eqnarray}
\label{Lambdac1} N_{\Lambda_{c}({\rm cud})}^{\rm coal}&\simeq&
g_{\Lambda_{c}({\rm cud})} N_{\rm c}N_{\rm u}N_{\rm d}
\prod_{i=1}^2\frac{(4\pi\sigma_i^2)^{3/2}}{V_{\rm
C}(1+2\mu_{i}T_{\rm C}\sigma_{i}^{2})}.
\end{eqnarray}
We note that the Wigner function of $\Lambda_c$  used in the above
does not take into account the $[ud]$ diquark correlation. This
correlation would reduce the width parameter for the relative wave
function of $u$ and $d$ quarks. Because the number of produced
$\Lambda_c$ is proportional to the third power of the width
parameter, treating $u$ and $d$ as independent quarks in the
$\Lambda_c$ thus gives an upper bound for the yield of $\Lambda_c$
from the coalescence of three independent $c$, $u$ and $d$ quarks in
QGP.

The contribution from coalescence of $c$ quarks with $[ud]$ diquarks
in QGP to the number of $\Lambda_{c}$ can be obtained by setting
$n=2$ in Eq.\ (\ref{coalmodel}), and replacing the Wigner functionof
$\Lambda_{c}$ by $\label{wigner2} f_{\Lambda_{c}({\rm c[ud]})}^{\rm
W}(x;p)= 8\exp (-{\mathbf{y}^{2}}/{\sigma_{{\rm c[ud]}}^{2}}
-\mathbf{k}^{2}\sigma_{{\rm c[ud]}}^{2}),$ where $\mathbf{y}$ and
$\mathbf{k}$ are the relative coordinate and momentum for the
two-body $c[ud]$ system, and $\sigma_{{\rm c[ud]}}=1/\sqrt{\mu_{{\rm
c[ud]}}\omega}$ with $\mu_{{\rm c[ud]}}=m_{\rm c}m_{\rm
[ud]}/(m_{\rm c}+m_{\rm [ud]})$. Then the result is
\begin{eqnarray}\label{lambda}
\label{lambdac2} N_{\Lambda_{c}({\rm c [ud]})}^{\rm coal}
\hspace*{-0.5em}&\simeq&\hspace*{-0.5em} g_{\Lambda_{c}({\rm c
[ud]})} N_{\rm c}N_{\rm [ud]} \frac{(4\pi\sigma_{{\rm
c[ud]}}^2)^{3/2}}{V_{\rm C}(1+2\mu_{{\rm c[ud]}} T_{\rm
C}\sigma_{{\rm c[ud]}}^{2})}
\end{eqnarray}
with $g_{\Lambda_{c}({\rm c[ud]})}=2 \times 1/3^2 \times 1/2 = 1/9$.
Contrary to the coalescence of independent $c$, $u$ and $d$ quarks
from QGP, where the $[ud]$ diquark substructure of $\Lambda_c$ is
neglected, it is here considered as a single entity as assumed for
the $[ud]$ diquark in QGP. The effect of finite structure of the
$[ud]$ diquark in both QGP and $\Lambda_c$ is expected to reduce the
yield of $\Lambda_c$ in comparison to that obtained from
Eq.~(\ref{lambdac2}). The latter thus also gives an upper bound for
$\Lambda_c$ production in the diquark picture.

The total yield of $\Lambda_{c}$ is given by the sum of above two
contributions, i.e., $N_{\Lambda_{c}}^{\rm coal} =
N_{\Lambda_{c}({\rm cud})}^{\rm coal} + N_{\Lambda_{c}({\rm c
[ud]})}^{\rm coal}.$ The $\Lambda_{c}$ yield can be compared with
the $D$ meson yield, which is not affected by the $ud$ diquark
correlation and is determined by an equation similar to
Eq.(\ref{lambda}) using instead the statistical factor $g_{\rm
D^{0}}=1 \times 1/3^2 \times 1/2^2 = 1/36$, the $u$ quark number
$N_{\rm u}$, the reduced mass $\mu_{\rm cu}=m_{\rm c} m_{\rm
u}/(m_{\rm c}+m_{\rm u})$, and the oscillator constant $\sigma_{\rm
cu}=1/\sqrt{\mu_{\rm cu}\omega}$.

\begin{figure}[tb]
\begin{center}
\includegraphics[height=2.0in,keepaspectratio,angle=0]{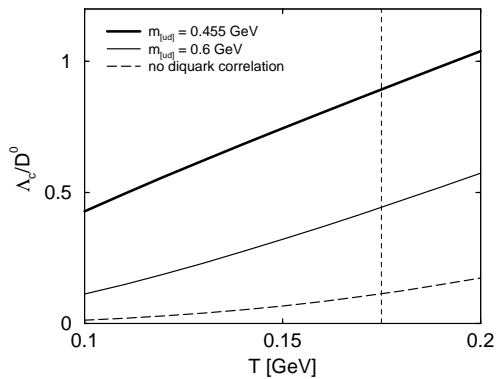}
\end{center}
\caption{The yield ratios $\Lambda_{c}/D^{0}$ as functions of
temperature\protect.} \label{number_ratio}
\end{figure}

Using the oscillator frequency $\omega=0.3$ GeV, determined from the
size $\langle r_{D_s}^2\rangle_{\rm ch}\simeq 0.124~{\rm fm}^2$ of
the $D_s^+(c\bar s)$ meson based on the light-front quark model
\cite{Hwang:2001th}, the resulting yield ratio $N_{\Lambda_{c}}^{\rm
coal}/N_{\rm D^{0}}^{\rm coal}$ ($\Lambda_{c}/D^{0}$), which is the
same in heavy ion collisions at RHIC and LHC, is plotted in
Fig.~\ref{number_ratio} as a function of the hadronization
temperature. It is seen that the yield ratio $\Lambda_{c}/D^{0}$ at
$T_{\rm C}=0.175$ GeV is $\simeq 0.11$ without diquarks in QGP (dashed line), and
it increases to $\simeq 0.44$ in the presence of the diquark $[ud]$
with mass $m_{\rm[ud]}=0.6$ GeV (thin solid line), corresponding to
a loosely bound state which can hardly exist near $T_{\rm C}$. If
the diquark mass has the minimum value $m_{\rm[ud]}=0.455$ GeV (bold
solid line), the $\Lambda_{c}/D^{0}$ ratio becomes even larger and
has a value of about 0.89 at $T_{\rm C}$. Therefore, the diquarks in
QGP raise the $\Lambda_{c}/D^{0}$ ratio by about a factor 4 - 8 in
comparison with the case without diquarks. For a QGP with a finite
baryon chemical potential, there are more quarks than antiquarks and
this would lead to a reduction of the $D^{0}$ yield and an increase
of the $\Lambda_{c}/D^{0}$ ratio. The expected enhancement of
$\Lambda_{c}/D^{0}$ ratio is still seen if we change the size of
$\Lambda_c$ by a reasonable amount. Because the enhanced
$\Lambda_{c}/D^{0}$ ratio is robust with respect to changes in the
binding energy of the diquark and the size of $\Lambda_c$, more
sophisticated models are not expected to modify our conclusion.

>From a simple application of the statistical hadronization model
\cite{PBM07}, the yield ratio $\Lambda_{c}/D^{0}$ is roughly
estimated as
$2(m_{\Lambda_c}/m_{D_0})^{3/2}\exp(-(m_{\Lambda_{c}}-m_{D^0})/T_{\rm
C}) \simeq 0.24$ at the hadronization temperature of $T_{\rm
C}=0.175$ GeV \cite{Lee:2007tn}. Although this value is a factor of
two larger than that from the above three-body coalescence of
independent $c$, $u$, and $d$ quarks, it is smaller than the case
that includes diquarks. A smaller production ratio is also observed
in elementary processes. In p+p collisions, 1630 $\Lambda_{c}$'s and
10210 $D^{0}$'s have been measured by SELEX at Fermi Lab, and this
gives a yield ratio $\Lambda_{c}/D^{0}\simeq 0.159$
\cite{Kushnirenko:2000ed}. In inclusive decay processes of a $B$
meson, the ratios are is $\Lambda_{c}/D^{0} \simeq 0.03$ and
$\Lambda_{c}/D^{-} \simeq 0.14$ from the measured fractions: 79\% of
$\bar{D}^{0}X$ and 2.8\% of $\bar{\Lambda}_{c}^{-}X$ in the $B^{+}$
decay, and 36.9\% of $D^{-}X$  and 5.0 \% of
$\bar{\Lambda}_{c}^{-}X$  in the $B^{0}$ decay, with arbitrary
hadrons $X$ \cite{Yao:2006px}. Since these experimental ratios
include $\Lambda_c$ and $D^0$ from decays of charmed resonances, a
more quantitative comparison requires the inclusion of the
resonances contribution in both the statistical and the coalescence
model \cite{yasui}.

The $\Lambda_{c}$ produced in QGP may change into a $D$ meson in the
hadronic phase due to collisions such as $\Lambda_{c} \pi
\rightarrow N D ( D^{\ast})$. With the pion threshold momentum
$p_{\rm th} \simeq 0.43$ GeV in the $\Lambda_{c} \pi \rightarrow N
D$ process, which is larger than the typical energy scale $T_{\rm
C}$, the conversion time due to this process is estimated as $1/\tau
= \Gamma_{\rm th} = 3 \int_{p_{\rm th}}^{\infty} \sigma n(p) {{\rm
d}^{3}p}/{(2\pi)^{3}},$ with 3 being the isospin factor, $\sigma$
the cross section, and $n(p)\simeq
\exp(-\sqrt{p^{2}+m_{\pi}^{2}}/T)$. With $\sigma=5$ mb as a
reasonable value suggested from the $J/\psi$ dissociation
\cite{Liu:2001ce} and $T=T_{\rm C}$ for simplicity, we obtain $\tau
\simeq 17.8$ fm, which is comparable with the lifetime of the
hadronic phase $t_{\rm H}\simeq 10$ fm \cite{Chen:2003tn}, leading
to a suppression factor $e^{-t_{\rm H}/\tau} \simeq 0.57$ for the
$\Lambda_{c}$ yield. Since the temperature in the hadronic phase is
lower than $T_{\rm C}$, the actual suppression factor will be closer
to one. Therefore, the $\Lambda_{c}$ enhancement is expected to
survive the hadronic processes.

In summary, assuming the existence of stable bound diquarks in the
strongly coupled QGP, we have discussed the enhancement of the
$\Lambda_{c}$ yield in heavy ion collisions, which is induced by the
two-body collision between the $c$ quark and the $[ud]$ diquark. The
$\Lambda_{c}$ enhancement would open a new way to find the existence
of QGP in heavy ion collisions and also provide an experimental tool
to probe the diquark correlation in QGP. This would, in turn,
confirm the diquark structure in heavy baryons with a single heavy
quark. It is interesting to note that the observed suppression of
the $D$ meson yield at RHIC \cite{Adler:2005ab,Laue:2004tf} could be
partially a consequence of the enhanced production of $\Lambda_{c}$
\cite{sorenson}.

Our study can be straightforwardly extended to $\Lambda_b$
production. Using the bottom quark production cross sections
predicted from the perturbative QCD for p+p collisions at RHIC \cite{vogt1} and
LHC \cite{vogt2}, we estimate the bottom quark numbers in one unit of midrapidity for
corresponding heavy ion collisions to be $\sim 0.02$ and $\sim 0.8$,
respectively. This leads to $\Lambda_b/B^0$ ratios of $0.098$,
$0.38$ and $0.82$ for the three scenarios of no diquark correlation
and diquark masses of $0.6$ and $0.45$ GeV, respectively. As in the
case of $\Lambda_{c}$, the diquark correlation gives rise to a large
enhancement in the $\Lambda_b/B_0$ ratio in heavy ion collisions.
Although the yield of $\Lambda_b$ in these collisions is much
smaller than that of $\Lambda_c$, its enhancement is a better signal
to be detected for QGP as the diquark picture would be more valid
than in $\Lambda_{c}$, and its much longer lifetime ($\tau \simeq
372 \mu$m) than that of $\Lambda_c$ ($\tau \simeq 62 \mu$m) will
also facilitate its detection. To study the enhancement of
$\Lambda_c$ and $\Lambda_b$ production is thus an interesting and
challenging subject at RHIC and LHC.

We thank Yongseok Oh for helpful discussions. The work was supported
by the Korea Research Foundation KRF-2006-C00011 and KRF-2007-313-C00175,
 and the US National
Science Foundation under Grant No.\ PHY-0457265 and the Welch
Foundation under Grant No.\ A-1358.


\end{document}